\documentclass[10pt]{article}
\usepackage{inputenc}
\usepackage{amsmath, amssymb, amsfonts}
\usepackage{graphicx,epstopdf,bm}
\usepackage{subfigure}
\usepackage[colorlinks,bookmarks,urlcolor=blue,citecolor=blue,linkcolor=blue]{hyperref}
\usepackage[T1]{fontenc}
\usepackage{xspace}
\usepackage[paperwidth=210mm, paperheight=297mm, left=100pt, top=60pt, textwidth=400pt, textheight=692pt, footskip=60pt]{geometry}
\usepackage{siunitx}

\newcommand{\vx}{\mathbf{x}}

\title{Convolutional neural networks automate detection for tracking of submicron scale particles in 2D and 3D}

\author{Jay M. Newby\thanks{Department of Mathematics, University of Alberta, Edmonton, AB, Canada, T6G 2R3} \and Alison M. Schaefer\thanks{Division of Pharmacoengineering and Molecular Pharmaceutics, Eshelman School of Pharmacy, University of North Carolina--Chapel Hill, Chapel Hill, NC 27599} \and Phoebe T. Lee\thanks{UNC-NCSU Joint Department of Biomedical Engineering, University of North Carolina--Chapel Hill, Chapel Hill, NC 27599} \and M. Gregory Forest\thanks{Department of Mathematics and Applied Physical Sciences, University of North Carolina--Chapel Hill, Chapel Hill, NC 27599} \and Samuel K. Lai\footnotemark[2]}

\begin{document}
\maketitle

\begin{abstract}  
Particle tracking is a powerful biophysical tool that requires conversion of large video files into position time series, i.e. traces of the species of interest for data analysis.  Current tracking methods, based on a limited set of input parameters to identify bright objects, are ill-equipped to handle the spectrum of spatiotemporal heterogeneity and poor signal-to-noise ratios typically presented by submicron species in complex biological environments.  Extensive user involvement is frequently necessary to optimize and execute tracking methods, which is not only inefficient but introduces user bias.  To develop a fully automated tracking method, we developed a convolutional neural network for particle localization from image data, comprised of over 6,000 parameters, and employed machine learning techniques to train the network on a diverse portfolio of video conditions.  The neural network tracker provides unprecedented automation and accuracy, with exceptionally low false positive and false negative rates on both 2D and 3D simulated videos and 2D experimental videos of difficult-to-track species.
\end{abstract}

\section{Significance Statement}
The increasing availability of powerful light microscopes capable of collecting terabytes of high-resolution 2D and 3D videos in a single day has created a great demand for automated image analysis tools. Tracking the movement of nanometer scale particles (e.g., virus, proteins, synthetic drug particles) is critical for understanding how pathogens breach mucosal barriers and for the design of new drug therapies. Our advancement is to use an artificial neural network that provides, first and foremost, substantially improved automation. Additionally, our method improves accuracy compared to current methods and reproducibility across users and labs.

\section{Introduction}
 
In particle tracking experiments, high-fidelity tracking of an ensemble of species recorded by high-resolution video microscopy can reveal critical information about species transport within cells or mechanical and structural properties of the surrounding environment. 
For instance, particle tracking has been extensively used to measure the real-time penetration of pathogens across physiological barriers \cite{wang2014,wang2016}, to facilitate the development of nanoparticle systems for transmucosal drug delivery \cite{lai2007rapid,yang2011}, to explore dynamics and organization of domains of chromosomal DNA in the nucleus of living cells \cite{vasquez2016}, and to characterize the micro- and meso- scale rheology of complex fluids via engineered probes \cite{mason1997,wirtz2009,chen2003,wong2004,waigh2005,flores2011,schultz2012,josephson2016,valentine2001,lai2010pnas}.  

There has been significant progress towards the goal of fully automated tracking, and dozens of methods are currently available that can automatically process videos, given a predefined set of adjustable parameters \cite{crocker1996,chenouard2014}.
The extraction of individual traces from raw videos is generally divided into two steps: (i) identifying the precise locations of particle centers from each frame of the video, and (ii) linking these particle centers across sequential frames into tracks or paths.  
Previous methods for particle tracking have focused more on the linking portion of the particle tracking problem.
Much less progress had been made on localization, in part because of the prevailing view that linking is more crucial, having the potential to correctly pick the true positives from a large set of localizations that may contain a sizable fraction of false positives.
In this paper, we primarily focus on localization instead of linking.
We present a particle tracking algorithm, constructed from a new localization algorithm and one of the simplest known linking algorithms, slightly modified from its most common implementation.

\begin{figure*}[htbp]
  \centering
  \includegraphics[width=14cm]{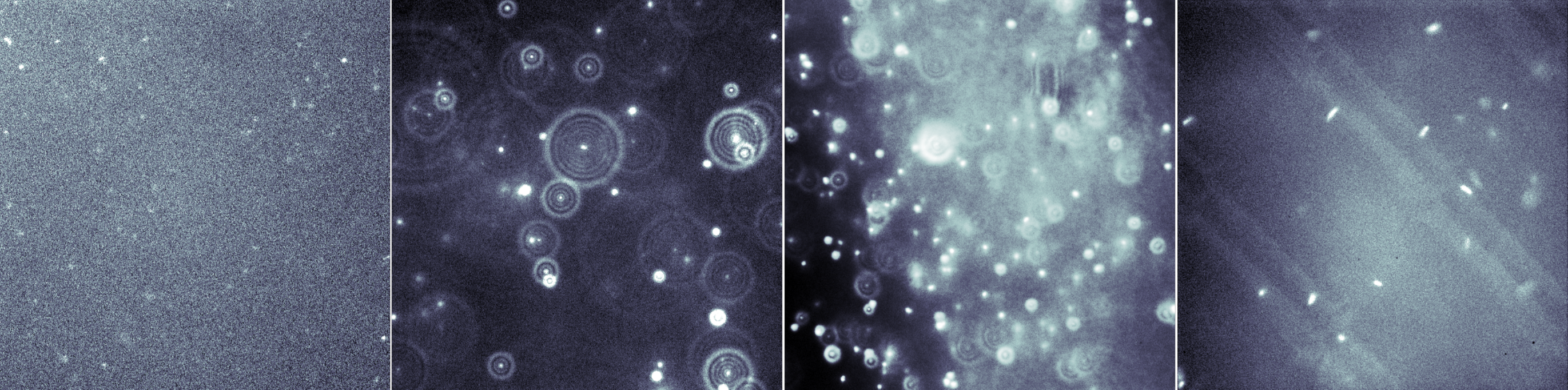}
  \caption{{\bf Sample frames from experimental videos, highlighting some of the challenging conditions for particle tracking.} (from left to right) 50nm particles captured at low SNR, 200nm particles with diffraction disc patterns, variable background intensity, and ellipsoid PSF shapes from $1-2\mu m$ Salmonella.}
  \label{fig:frames}
\end{figure*}

\begin{figure}[htb]
\centering
\includegraphics[width=8cm]{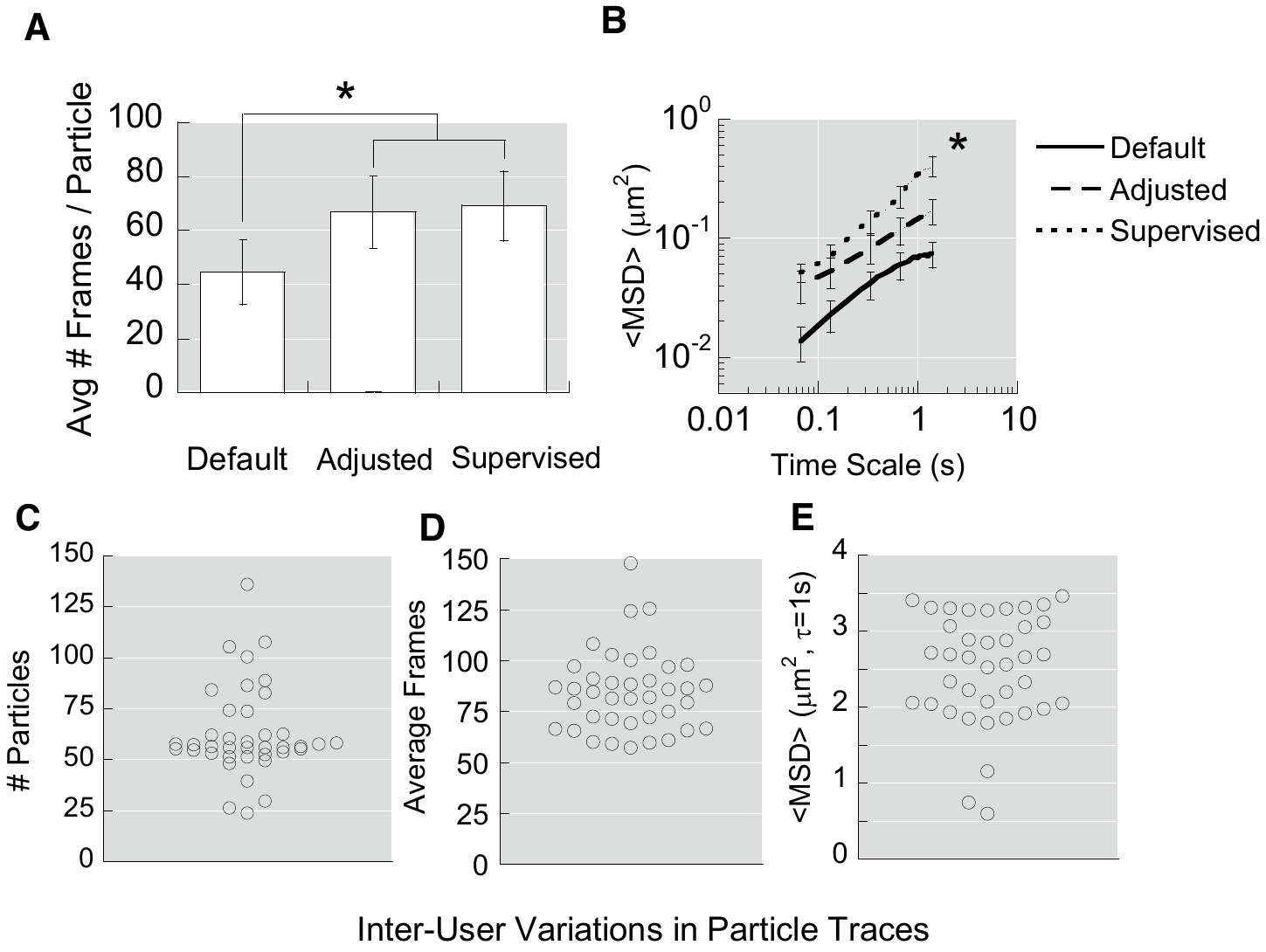}
  \caption{{\bf The need for supervision in particle tracking, and inter-user variations in supervised tracking data.} Data represents the average of 4 movies of muco-inert 200 nm PEGylated polystyrene beads in human cervicovaginal mucus.  Data from human supervised tracking (Supervised), which includes manually inspecting paths to remove false positives and minimize false negatives, is compared to results generated under default conditions of the tracking software (Default) and conditions manually adjusted by the user (Adj) to improve tracking accuracy.  (A) average frames per particle; (B) ensemble-averaged geometric mean square displacements ($\langle {\rm MSD}\rangle$) vs. time scale.  Error bars represent standard error of the mean.  * indicates statistically significant difference compared to `Standard' ($p < 0.05$).  (C-E) Inter-user variations in particle tracking.  Different tracking software users were asked to analyze the same video of 200 nm bead in human cervicovaginal mucus. (A) Total particles tracked; (B) average frames tracked per particle; and (C) ensemble-averaged geometric mean square displacements ($\langle {\rm MSD}\rangle$) at a time scale ($\tau$) of 1s.}
\label{fig:PTcompare}
\end{figure}

The primary novelty of our method is automation and accuracy.
Even though many particle tracking methods have been developed that can automatically process videos, when presented with videos containing spatiotemporal heterogeneity (see Fig.~\ref{fig:frames}) such as variable background intensity, photobleaching or low signal-to-noise ratio (SNR), the set of parameters used by a given method must be optimized for each set of video conditions, or even each video, which is highly subjective in the absence of ground truth.  
Parameter optimization is time consuming and requires substantial user guidance.  
Furthermore, when applied to experimental videos, user input is still frequently needed to remove phantom traces (false positives) or add missing traces (false negatives) (Fig.~\ref{fig:PTcompare}A-B).  
Thus, instead of providing full automation, current software is perhaps better characterized as facilitating supervised particle tracking, requiring substantial human interaction that is time consuming and costly. 
More importantly, the results can be highly variable, even for the same input video (Fig.~\ref{fig:PTcompare}C-E).

A major dificulty for optimizing tracking methods for specific experimental conditions is access to ``ground truth,'' which can be highly subjective and labor intensive to obtain.
One approach for applying a tracking method to experimental videos is to tune parameter values by hand, while qualitatively assessing error accross a range of videos.
This proceedure is laborious and subjective.
A better approach, using quantitative optimization, is to generate simulated videos---for which ground truth is known---that match as closely as possible to the observed experimental conditions. 
Then, a given tracking method suitable for those conditions can be applied to the simulated videos, and the error quantitatively assessed. 
By quantifying the tracking error, the parameters in the tracking method can be systematically optimized to minimize the tracking error over a large number of videos.  
Finally, once the parameters have been optimized on simulated data, the same parameters can be used (after fine tuning parameters and adding or removing traces to ensure accuracy) to analyze experimental videos.

To overcome the need to optimize parameters for each video condition, we take the aforementioned methodology to the next logical step: instead of optimizing for a specific microscopy conditions, we compile a large portfolio of simulations that encompasses the wide spectrum of potential variations encountered in particle tracking experiments.   
Existing methods are designed with as few parameters as possible to make the software simple to use, and a single set of parameters can usually be found for a specific microscopy conditions (SNR, size, shape, etc.) that identifies objects of interest.
Nevertheless, a limited parameter space compromises the ability to optimize the method for a large portfolio of conditions. 
An alternative approach is to construct an algorithm with thousands of parameters, and employ machine learning to optimize the algorithm to perform well under all conditions represented in the portfolio. 
Here, we adapt an existing neural network imaging framework, called a convolutional neural network (CNN), to the challenge of particle identification---which is a novel application for CNN-type neural networks.

CNNs have become the state-of-the-art for object recognition in computer vision, outperforming other methods for many imaging tasks \cite{krizhevsky2012,long2015}. 
A CNN is a type of feed-forward artificial neural network designed to process information in a layered network of connections.
The linking stage of particle tracking is sometimes viewed as the most critical for accuracy.
Here, we develop a novel approach for particle identification, while using one of the simplest particle linking strategies, namely, adaptive linear asignment \cite{jaqaman2008}.
We rigorously test the accuracy of our method, and find substantial improvement (in terms of false positives and false negatives) over several existing methods, suggesting that particle identification is the most critical component of a particle tracking algorithm, particularly for automation.

A number of research groups are beginning to apply machine learning to particle tracking \cite{boland2001,jiang2007,smal2010}, primarily involving `hand crafted' features that in essence serve as a set of filter banks for making statistical measurements of an image, such as mean intensity, standard deviation and cross correlation.  These features are used as inputs for a support vector machine, which is then trained using machine learning methods. The use of hand-crafted features substantially reduces the number of parameters that must be trained. 

In contrast, we have developed our network to be trained end-to-end, or pixels-to-pixels, so that the input is the raw imaging data, and the output is a probabilistic classification of particle versus background at every pixel.
Importantly, we have designed our network to be ``recurrent'' in time so that past and future observations are used to predict particle locations.

In this paper, we construct a CNN, comprised of a 3-layer architecture and over 6,000 tunable parameters, for particle localization.  
All of the neural network's tunable parameters are optimized using machine learning techniques, which means there are never any parameters that the user needs to adjust for particle localization. 
The result is a highly optimized network that can perform under a wide range of conditions without any user supervision.
To demonstrate accuracy, we test the neural network tracker on a large set of challenging videos that span a wide range of conditions, including variable background, particle motion, particle size, and low SNR

\section{Simulation of 4D greyscale image data}
\label{sec:vids}
To train the network on a wide range of video conditions, we developed new video simulation software that accounts for a large range of conditions found in particle tracking videos (see Fig.~\ref{fig:frames}).
The primary advance is to include simulations of how particles moving in 3D appear in a 2D image slice captured by the camera. 

A standard camera produces images that are typically single channel (grey scale), and the image data is collected into four dimensional (three space and one time dimension) arrays of 16 bit integers.
The resolution in the $(x,y)$ plane is dictated by the camera and can be in the megapixel range.
The resolution in the $z$ coordinate is much smaller since each z-axis slice imaged by the camera requires a piezo-electric motor to move the lense relative to the sample.
A good piezo-electric motor is capable of moving between z-axis slices within a few milliseconds, which means that there is a trade off between more z-slices and the over all frame rate.
For particle tracking, a typical video includes 10-50 z-slices per volume.
The length of the video refers to the number of time points, i.e., the number of volumes collected.
Video length is often limited by photobleaching, which slowly lowers the SNR as the video progresses.

To simulate a particle tracking video, we must first specify how particles appear in an image. 
We refer to the pixel intensities captured by a microscope and camera resulting from a particle centered at a given position $(x,y,z)$ as  the observed point spread function (PSF), denoted by $\psi_{ijk} (x, y, z)$, where $i,j,k$ are the pixel indices. 
The PSF becomes dimmer and less focused as the particle moves away from the plane of focus $(z=0)$. 
Away from the plane of focus, the PSF also develops disc patterns caused by diffraction, which can be worsened by spherical aberration. 
While deconvolution can mitigate the disc patterns appearing in the PSF, the precise shape of the PSF must be known or unpredictable artifacts may be introduced into the image.

The shape of the PSF depends on several parameters that vary depending on the microscope and camera, including emitted light wavelength, numerical aperture, pixel size, and the separation between z-axis slices. 
While physical models based on optical physics that expose these parameters have been developed for colloidal spheres \cite{bierbaum2017light}, it is not practical for the purpose of automated particle tracking within complex biological environments.
In practice, there are many additional factors that affect the PSF, such as the refractive index of the glass slide, of the lens oil (if oil-immersion objective is used), and of the medium containing the particles being imaged. 
The latter presents the greatest difficulty since biological specimens are often heterogeneous, and their optical properties are difficult to predict. 
The PSF can also be affected by particle velocity, depending on the duration of the exposure interval used by the camera. 
This makes machine learning particularly appealing, because we can simply randomize the shape of the PSF to cover a wide range of conditions, and the resulting CNN is capable of automatically `deconvolving' PSFs without the need to know any of the aforementioned parameters.

Low SNR is an additional challenge for tracking of submicron size particles.
High performance digital cameras are used to record images at a sufficiently high frame rate to resolve statistical features of particle motion.
Particles with a hydrodynamic radius in the range of 10-100nm move quickly, requiring a small exposure time to minimize dynamic localization error (motion blur) \cite{savin2005}.
Smaller particles also emit less light for the camera to collect.
To train the neural network to perform in these conditions, we add Poisson shot noise with random intensity to the training videos.
We also add slowly varying random background patterns (see Supplemental Figure 1).

\section{An artificial neural network for particle localization}
\begin{figure}[htb]
  \centering
  \includegraphics[width=6cm]{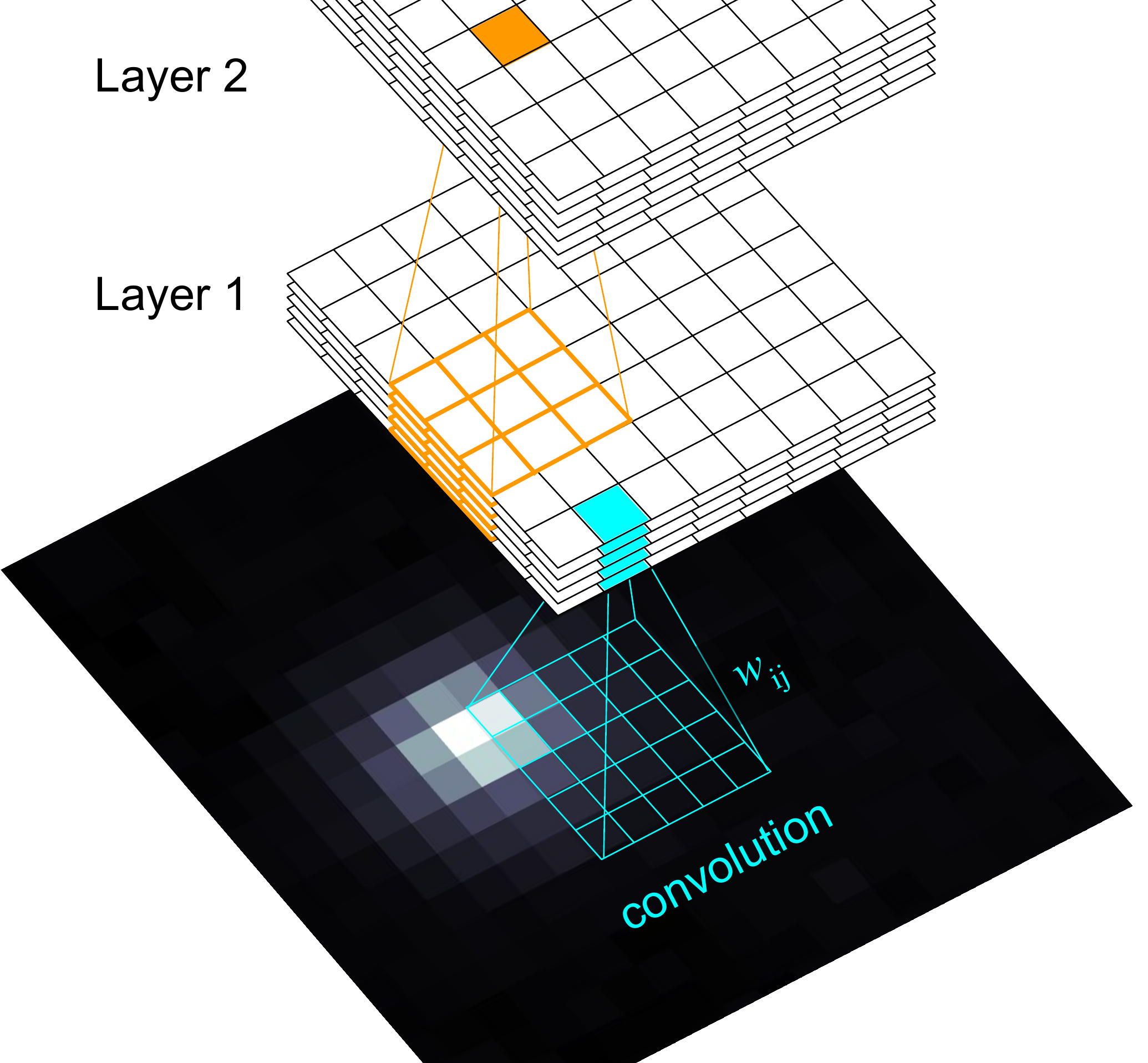}
\caption{{\bf The convolutional neural network.}  Diagram of the layered connectivity of the artificial neural network.}
  \label{fig:cnn}
\end{figure}

The `neurons' of the artificial neural network are arranged in layers, which opperate on multi-dimensional arrays of data.
Each layer output is 3 dimensional, with 2 spatial dimensions and an additional `feature' dimension (see Fig.~\ref{fig:cnn}).
Each feature within a layer is tuned to respond to specific patterns, and the ensemble of features are sampled as input to the next layer to form features that recognize more complex patterns.
For example, the lowest layer is comprised of features that detect edges of varying orientation, and the second layer features are tuned to recognize curved lines and circular shapes.

Each neuron in the network processes information from spatially local inputs (either pixels of the input image or lower layer neurons).
This enables a neuron to, figuratively speaking, see a local patch of the input image, which is smaller than the entire input image. 
The size of the image patch that affects the input to a given neuron is called its receptive field.
The input and output, denoted by $I_{ij}$ and $O_{ij}$, relationship for each neuron is given by $O_{ij} = F(\sum_{i',j'}w_{i',j'}I_{i+i',j+j'} - b)$, where the kernel weights $w_{ij}$ and output bias $b$ are trainable parameters.
Each layer has its own set of biases, one for each feature, and each feature has its own set of kernel weights, one for each feature in the layer directly below.
The nonlinearity $F(\cdot)$ is a pre-specified function that determines the degree of `activation' or output, we use $F(u) = \log(e^{u} + 1)$.
Inserting nonlinearity in between each layer of neurons is necessary for CNNs to robustly approximate nonlinear functions.
The most common choice is called the rectified linear unit ($F(u\geq 0) = u$ and $F(u < 0) = 0$).
Instead, we use a function with a similar shape that is also continuously differentiable, which helps minimize training iterations where the model is stuck in local minima \cite{glorot2011}.

The neural network is comprised of three layers; 12 features in layer one, 32 features in layer two, and the final two output features in layer three.
The output of the neural net, denoted by $q_{ijk}$, can be interpreted as the probability of a particle centered at pixel $(i,j,k)$.
We refer to these as detection probabilities.

While it is possible to construct a network that takes 3D image data as input, it is not computationally efficient.
Instead, the network is designed to process a single 2D image slice at a time (so that it can also be applied to the large set of existing 2D imaging data) while still maintaining the ability to perform 3D tracking.
Constructing 3D output $q_{ijk}$ is achieved by applying the network to each z-axis slice of the input image, the same way a microscope obtains 3D images by sequentially capturing each z-axis slice.
Two or three dimensional paths can then be reconstructed from the network output as described below.

We designed our network to be ``recurrent'' in time so that past and future observations are used to predict particle locations.
In particular, we use the forward-backward algorithm \cite{rabiner1986introduction} to improve accuracy.
Because detections include information from the past and future, the detection probabilities are reduced when a particle is not detected in the previous frame (the particle just appeared in the current frame) or is not detected in the following frame (the particle is about to leave the plane of focus).
Below, we show how the detection probabilities can be used by the linking algorithm to improve its performance.

\subsubsection*{Optimizing the neural network parameters}
The values of the trainable parameters in the network, including the kernel weights and biases, are optimized through the process of learning.
Using known physical models of particle motion and imaging, we simulate random particle paths and image frames that cover a wide range of conditions, including particle point spread function shape, variable background, particle number, particle mobility, and SNR.
The `ground truth' for each image consists of a binary image with pixels values $p_{ijk} = 1$ if $\Vert (j,i,k) - \vx_{n}\Vert < 2$ and $p_{ijk}=0$ otherwise.
Each training image is processed by the neural net, and the corresponding output is compared to the ground truth using the cross entropy error:
\begin{equation}
  \label{eq:3}
  H[p, q] = -\frac{1}{N}\sum_{ijk}\left[p_{ijk}\log q_{ijk} + (1-p_{ijk})\log (1-q_{ijk})\right],
\end{equation}
where $N$ is the total number of pixels in the image.
Further details can be found in the Supplementary Material.

\subsection*{Particle path linking}
 
From the neural net output, we extract candidate particles along with their probabilities through thresholding the detection probabilities $q_{ijk}$, where $ijk$ are the indices for each pixel of a single video frame.
The threshold of $q = 0.5$ represents a maximum likelihood classification: everything above $q = 0.5$ represents pixels corresponding to the presence of a nearby particle, and everything below this threshold is most likely part of the image background.
The pixels above threshold are grouped into candidate particles using the method of connected components \cite{lumia1983}.
Connected sets of nearest neighbor pixels $\mathcal{P}_{n}$ above the threshold are collected as candidate particles.
That is, $\mathcal{P}_{n}$ is a connected set and $q_{ijk}\geq 0.5$ for all $q_{ijk}\in \mathcal{P}_{n}$.

Each candidate particle is assigned the largest detection probability from its constituent detection probabilities within the connected component, i.e., $\rho_{n}= \max \mathcal{P}_{n}$.
The position of each candidate particle is taken to be the center of mass given by, $\vx_{n} = \frac{\sum_{q_{ijk} \in \mathcal{P}_{n}} (j, i, k) q_{i j k}}{\sum_{q_{ijk} \in\mathcal{P}_{n}}p_{ijk}}$.
Note that there are alternative particle localization methods \cite{parthasarathy2012} that may increase accuracy.
We have found that the center of mass method yields consistent sub-pixel accuracy of 0.6 pixels on average, which is sufficient for tracking tasks that require high accuracy such as micro-rheology.
The next stage is to link candidate particles from one frame to the next.
 
The dynamics of particle motion can vary depending on the properties of the surrounding fluid and the presence of active forces (e.g., flagellar mediated swimming of bacteria and molecular motor cargo transport).
In order to reconstruct accurate paths from a wide range of movement characteristics, we develop a minimal model.
A minimal assumption for tracking is that the observation sequence approximates continuous motion of an object.
To accurately capture continuous motion sampled at discrete time intervals, dictated by the camera frame rate, the particle motion must be sufficiently small between image frames.
Hence, we assume only that particles move within a Gaussian range from one frame to the next.

Let $\mathcal{L}_{t}$ denote the set of linked particle pairs $(\vx_{t}, \vx_{t+1})$ together with their probabilities $(\rho_t, \rho_{t+1})$ in frame $t$ to $t+1$.  
We must also consider the possibility that a given particle has just entered or is about to leave the image.
Let $\mathcal{N}_{t}^{\pm}$ be the set of probabilities for particles in frame $t$ that are not linked to a particle in frame $t\pm 1$. 
Then, the log likelihood cost of the link assignments (or lack of assignment) from frame $t$ to frame $t+1$ is given by
\begin{equation}
  \label{eq:1}
  \begin{split}
L_{t} &= 
 -\sum_{\vx_{t}, \vx_{t+1}\in \mathcal{L}_{t}}\frac{\Vert \vx_{t} - \vx_{t+1}\Vert^2}{2\sigma^{2}}  \\
&\quad  +\sum_{\rho_{t}, \rho_{t+1}\in \mathcal{L}_{t}}\left[ \log \rho_{t} + \log \rho_{t+1}\right] \\
&\quad-\sum_{\rho_{t}\in \mathcal{N}_{t}^{+}} \log(1-\rho_{t}) - \sum_{\rho_{t+1}\in \mathcal{N}_{t+1}^{-}} \log(1-\rho_{t+1}).
\end{split}
\end{equation}
The standard deviation $\sigma$ is a user-specified parameter.
Maximization of \eqref{eq:1} can be formulated as a linear programming problem, which we solve using the Hungarian-Munkres algorithm \cite{jaqaman2008}. 
 
Note that we have made a slight modification to the adaptive linking method developed in \cite{jaqaman2008}, where we have made use of the detection probabilities.
The standard (non adaptive) approach is to assign a penalty for not assigning a link to a particle, based on a fixed cutoff distance. 
Our adaptive scheme uses the detection probabilities as a variable cost for not assigning a link to a particle. 
The lower the detection probability for a particle (due to faint signal or absence in past or future frames), the lower the cost of failing to assign it a link.

We note that $\sigma$ is the only parameter in our tracking method (there are no adjustable parameters in the neural network localizer).
It is reasonable to be concerned about automation when the method contains an adjustable parameter.
Because of the adaptive nature of our linking algorithm, which is armed with certainty estimates from the neural network, we have found that in practice, $\sigma$ rarely needs to be adjusted.
In fact, every one of the $\sim$600 videos tracked for testing purposes in this paper used the same value of this parameter ($\sigma=20$). In the future, it may be possible to eliminate this parameter completely using a more sophisticated linking algorithm, such as a 'particle filter' \cite{blake1997condensation}, which is a Bayesian framework that is compatible with the neural network. 
Moreover, noise in particle localizations can arise from many factors, including low SNR image conditions. Kalman filters have been applied to path linking to reconstruct more accurate paths from noisy localization  \cite{wu2010analysis}.

\subsection*{Performance evaluation and comparison to existing software}
\begin{figure*}[tb]
\centering
\includegraphics[width=14cm]{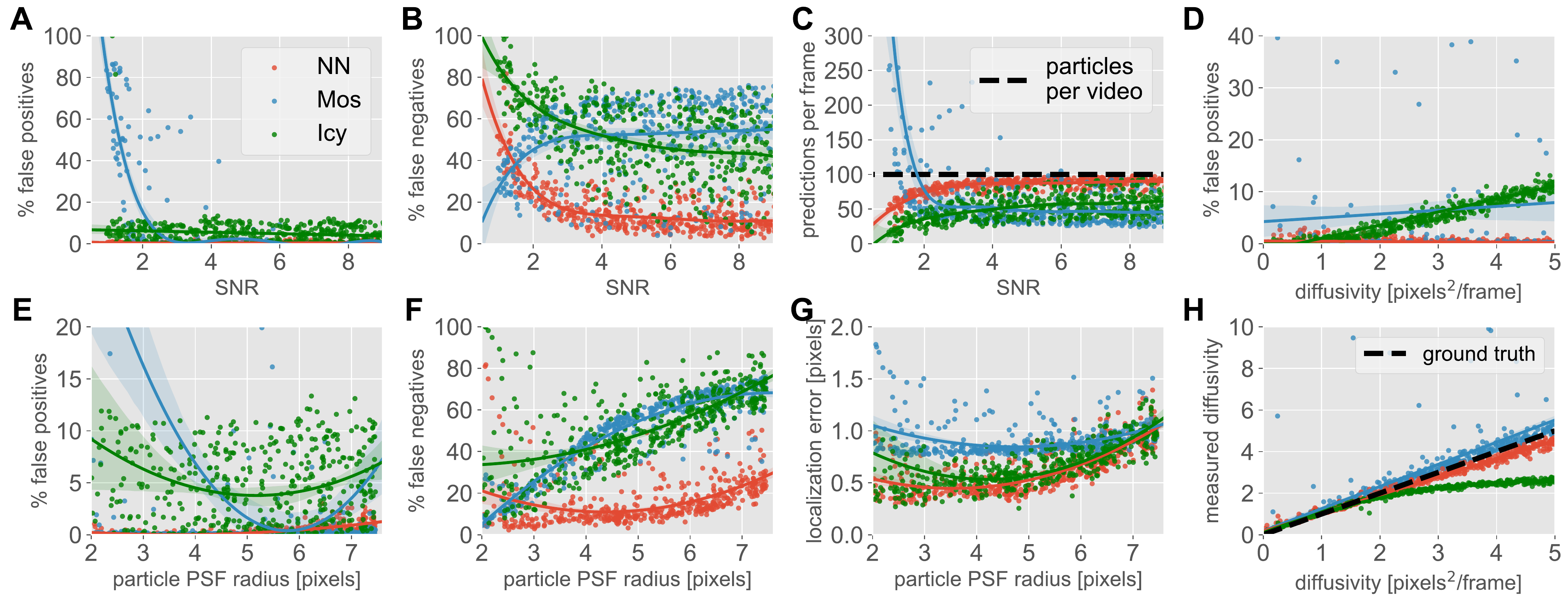}
\caption{{\bf Sensitivity analysis for randomized 2D synthetic test videos.} (A-C) 2D test results showing the (A) percentage of false positives, (B) percentage of false negatives and (C) predictions per frame vs. SNR.
 Mosaic shows a sharp rise in false positives for ${\rm SNR}<2$ (panel A), due to substantially more predictions than actual particles (panel C). Conversely, the neural net (NN) and Icy showed no increase in false positives at low SNR.  
(E-G)  results showing the (E) percentage of false positives, (F) percentage of false negatives and (G) localization error vs. the PSF radius.  (D-H) results showing the (D) percentage of false positives and (H) measured diffusivity vs. the ground truth particle diffusivity.}
\label{fig:synthTest}
\end{figure*}

We consider the primary goal for a high fidelity tracker to be accuracy (i.e., minimize false positives and localization error), followed by the secondary goal of maximizing data extraction (i.e., minimize false negatives and maximize path length).  
To gauge accuracy, particle positions were matched to ground truth using optimal linear assignment. 
The algorithm finds the closest match between tracked and ground truth particle positions that are within a preset distance of 5 pixels; this is well above the sub-pixel error threshold of 1 pixel, but sufficiently small to ensure 1-1 matching. 
Tracked particles that did not match any ground truth particles were deemed false positives, and ground truth particles that did not match a tracked particle were deemed false negatives. 
To assess the performance of the neural net tracker, we analyzed the same videos using three different leading tracking software that are publicly available:
\begin{itemize}
\item {\bf Mosaic (Mos)}: an ImageJ plug-in capable of automated tracking in 2D and 3D \cite{xiao2016} 
\item {\bf Icy}: an open source bio-imaging platform with pre-installed plugins capable of automated tracking in 2D and 3D \cite{olivo2002,chenouard2013}
\item {\bf Video Spot Tracker (VST)}: a stand-alone application developed by the Center for Computer-Integrated Systems for Microscopy and Manipulation at UNC-CH capable of 2D particle tracking. VST also has a convenient graphic user interface that allows a user to add or eliminate paths (because human-assisted tracking is time consuming, 100 2D videos were randomly selected from the 500 video set)
\end{itemize}

For the sake of visual illustration, we supplement our quantitative testing with a small sample of real and synthetic videos with localization indicators (see Supplementary Videos). 
In each video, red diamond centers indicate each localization from the neural network.

\subsubsection*{Performance on simulated 2D videos}

Because manual tracking by humans is subjective, our first standard for evaluating the performance of the neural net tracker (NN) and other publicly available software is to  test on simulated videos, for which the ground truth particle paths are known. 
The test included 500 2D videos and 50 3D videos, generated using the video simulation methodology described in Section \ref{sec:vids}.
Each 2D video contained 100 simulated particle paths for 50 frames at 512x512 resolution, (see Supplemental Figure 1). 
Each 3D video contained 20 evenly spaced z axis image slices of a 512x512x120 pixel region containing 300 particles. 
The conditions for each video were randomized, including variable background intensity, PSF radius (called particle radius for convenience), diffusivity, and SNR. 
Note that SNR is defined as the mean pixel intensity contributed by the particle PSFs divided by the standard deviation of the background pixel intensities.

To assess the robustness of each tracking method/software, we used the same set of tracker parameters for all videos (see Supplementary material for further details).  
Scatter plots of the 2D test video results for neural network tracker, Mosaic, and Icy are shown in Fig.~\ref{fig:synthTest}. 
For Mosaic, the false positive rate was generally quite low ($\sim$2\%) when ${\rm SNR} > 3$, but showed a marked increase to >20\% for ${\rm SNR} < 3$ (Fig.~\ref{fig:synthTest}A).  
The average false negative rates were in excess of 50\% across most ${\rm SNR} > 3$ (Fig.~\ref{fig:synthTest}B).  
In comparison, Icy possessed higher false positive rates than Mosaic at high SNR and lower false positive rates when SNR is decreased below 2.5, with a consistent $\sim$5\% false positive rate across all SNR values (Fig.~\ref{fig:synthTest}A).  
The false negative rates for Icy was greater than Mosaic at high SNR, and exceeded $\sim$40\% for all SNR tested (Fig.~\ref{fig:synthTest}B).  

All three methods showed some minor sensitivity in the false positive rate and localization error to the PSF radius (Fig.~\ref{fig:synthTest}E,G).
(Note that the high sensitivity Mosaic displayed to changes in SNR made the trend for PSF radius difficult to discern.)
Mosaic and Icy showed much higher sensitivity in the false negative rate to PSF radius, each extracting nearly 4-fold more particles as the PSF radius decreased from 8 to 2 pixels (Fig.~\ref{fig:synthTest}F).

One common method to analyze and compare particle tracking data is the ensemble mean squared displacement (MSD) calculated from particle traces. Since the simulated paths in the 2D and 3D test videos were all Brownian motion (with randomized diffusivity), we have that $\langle \vert{\bf x}(t)\vert^{2}\rangle = 4Dt$, where $D$ is the diffusivity. To make a simple MSD comparison for Brownian paths, we computed estimated diffusivities using the MSD at the path duration $1< T\leq 50$, with $D \approx \langle \vert{\bf x}(T)\vert^{2}\rangle/(4T)$.  (See Fig.\ref{fig:realTest} for an MSD analysis on experimental videos of particle motion in mucus.) When estimating diffusivities, Icy exhibited increased false positive rates with faster moving particles (Fig.~\ref{fig:synthTest}D), likely due to the linker compensating for errors made by the detection algorithm.  In other words, while the linker was able to correctly connect less-mobile particles without confusing them with nearby false detections, when the diffusivity rose, the particle displacements tended to be larger than the distance to the nearest false detection. Consequently when $D>2$, the increased false positives along with increased increment displacements caused Icy to underestimate the diffusivity (Fig. \ref{fig:synthTest}H) because paths increasingly incorporated false positives.

In contrast to Mosaic and Icy, the neural network tracker possessed a far lower mean false positive rate of $\sim$0.5\% across all SNR values tested (Fig.~\ref{fig:synthTest}A).  
The neural network tracker was able to achieve this level of accuracy while extracting a large number of paths, with <20\% false negative rate for all ${\rm SNR} > 2.5$ and only a modest increase in the false negative rate at lower SNR (Fig.~\ref{fig:synthTest}B). 
Importantly, the neural network tracker performed well under low SNR conditions by making fewer predictions, and the number of predictions made per frame are generally in reasonable agreement with the theoretical maximum (Fig.~\ref{fig:synthTest}C).  
Since the neural network was trained to recognize a wide range of PSFs, it also maintained excellent performance (<1\% false positive, <20\% false negative) across the range of PSF radius (Fig.~\ref{fig:synthTest}F).  
The neural network tracker possessed comparably good localization error as Mosaic and Ivy, less than one pixel on average and never more than two pixels, even though true positives were allowed to be as far as 5 pixels apart (Fig.~\ref{fig:synthTest}G).  

\subsubsection*{Performance on simulated 3D videos}

\begin{figure*}[htb]
\centering
\includegraphics[width=10cm]{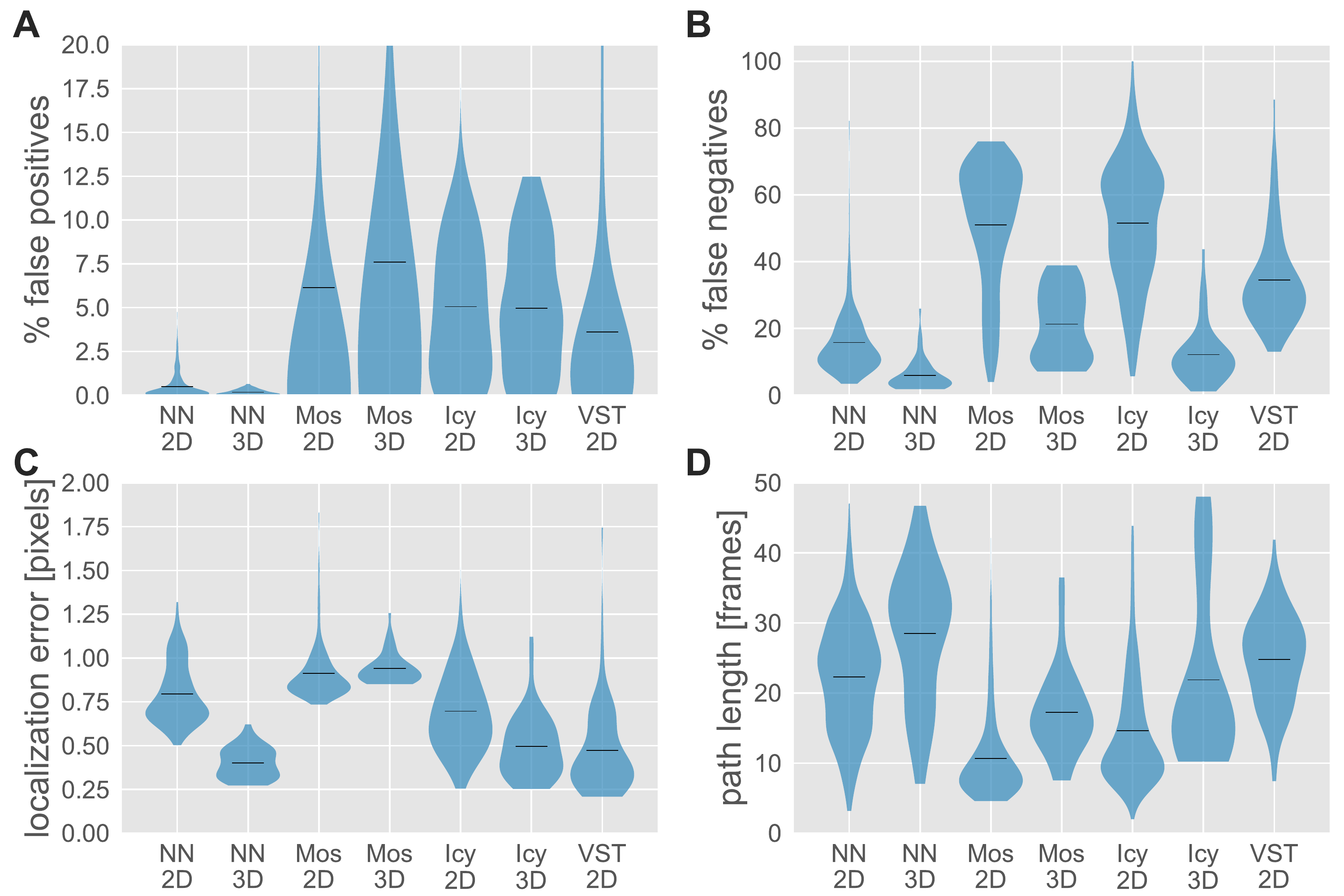}
\caption{{\bf Violin plots showing the performance on 3D test videos for each of the four methods: the neural network tracker (NN), Mosaic (Mos), Icy, and VST.} Performance with 2D simulated videos are included for comparison. The solid black lines show the mean, and the thickness of the filled regions show the shape of the histogram obtained from 500 (50) randomized 2D (3D) test videos. Note that the VST results only included 100 test videos. }
\label{fig:synthTestSumm}
\end{figure*}

When analyzing 3D videos, Mosaic and Icy were able to maintain roughly comparable false positive rates ($\sim$5-8\%) as analyzing 2D videos (Fig.~\ref{fig:synthTestSumm}A).  
Surprisingly, analyzing 3D videos with the neural network tracker resulted in an even lower false positive rate than 2D videos, with $\sim$0.2\% false positives.  
All three methods capable of 3D tracking exhibited substantial improvements in reducing false negatives, reducing localization error, and increasing path duration (see Fig.~\ref{fig:synthTestSumm}B-D).  
Strikingly, the neural network was able to correctly identify an average of $\sim$95\% of the simulated particles in a 3D video, i.e., <5\% false negatives, with the lowest localization error as well as the longest average path duration among the three methods.

\subsubsection*{Performance on experimental 2D videos}

\begin{figure*}[tbh]
  \centering
  \includegraphics[width=12cm]{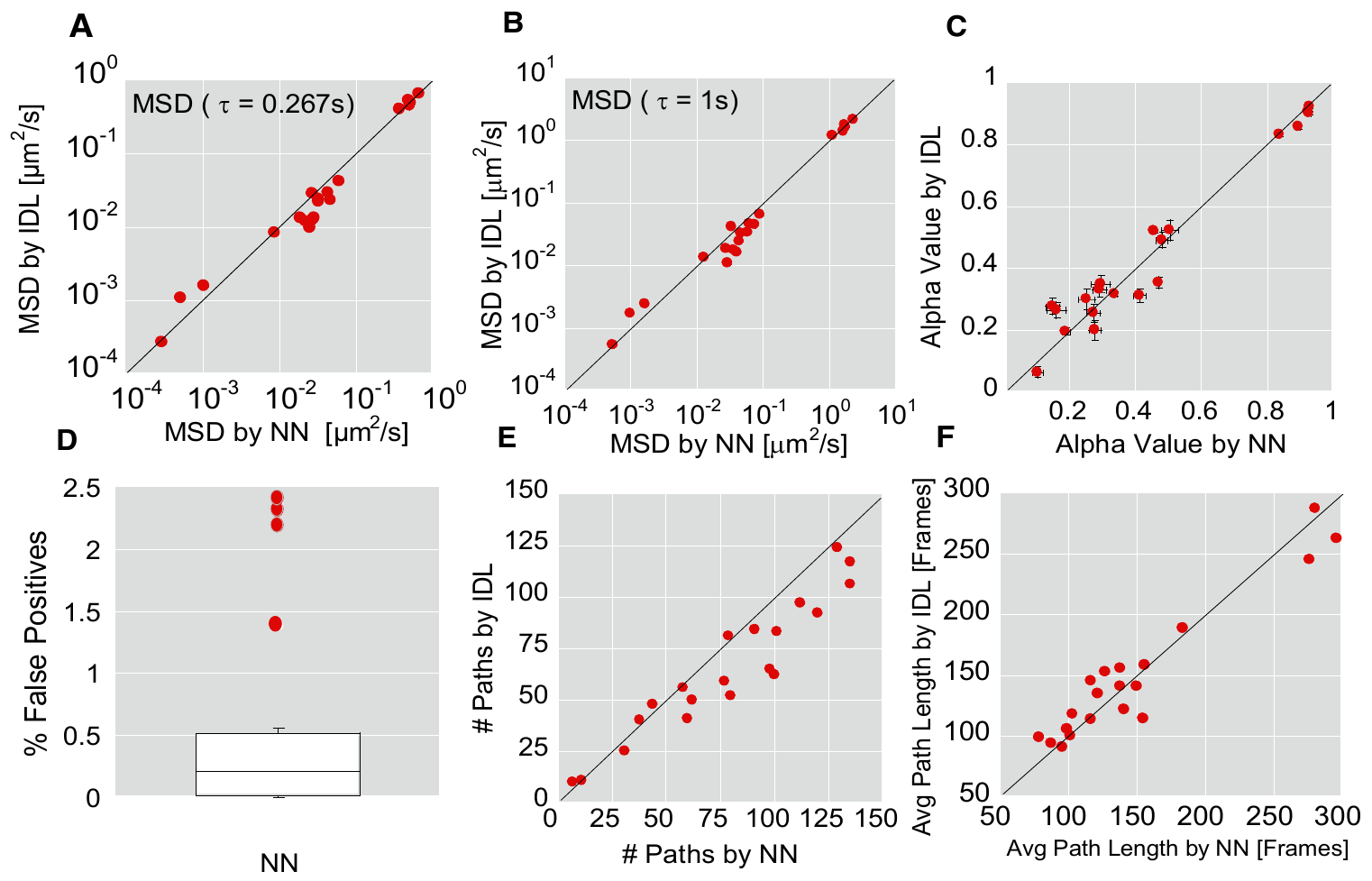}
  \caption{{\bf Comparison of human tracked (assisted by the commercially available software package IDL) and neural network tracked output.} 
Ensemble-averaged geometric mean square displacements ($\langle {\rm MSD}\rangle$) at a time scale ($\tau$) of (A) $0.267 s$ and (B) $1 s$. (C) alpha value (D) percentage of false positives normalized by path-length (E) number of particles tracked (F) average path duration per particle. The error bars in (C) represent standard error of the mean. The box plot in (D) shows symbols for the outliers above the 80th percentile of observations. The data set includes 20 different movies encompassing muco-inert 200 nm PEGylated polystyrene beads 200 nm carboxylated beads, HIV virus-like particles and herpes simplex virus in human cervicovaginal mucous. Further details regarding the experimental conditions for the videos used in the test can be found in the Methods Section.}
  \label{fig:realTest}
\end{figure*}

Finally, we sought to evaluate the performance and rigor of the neural network tracker on experimentally-derived rather than simulated videos, since the former can include spatiotemporal variations and features that might not be captured in simulated videos.  Because analysis from the particle traces can directly influence interpretations of important biological phenomenon, the common practice is for the end-user to supervise and visually inspect all traces to eliminate false positives and minimize false negatives.  Against such rigorously verified tracking, the neural net tracker was able to produce particle paths with comparable mean squared displacements across different time scales, alpha values, a low false positive rate, greater number of traces i.e. decrease in false negative, and comparable path length (see Fig.\ref{fig:realTest}).  Most importantly, these videos were processed in less than one twentieth of the time it took to manually verify them, generally taking 30-60 seconds to process a video compared to 10-20 minutes to verify accuracy.

\subsection*{Discussion}

Although tracking the motion of large, bright, micron-sized beads is straightforward, it remains exceptionally difficult to rapidly and accurately obtain traces of entities, such as ultrafine nanoparticles and viruses, that are sub-micron in size. Sub-micron particles can readily diffuse in and out of the plane of focus, possess low SNR or significant spatial heterogeneity, and undergo appreciable photo-bleaching over the timescale of imaging.  Accurate conversion of videos to particle paths for these entities necessitates extensive human intervention; it is not surprising to spend 10-20x more time on extracting path data from videos than the actual video acquisition time.  Worse, substantial user variations is common even when using the same software to analyze the same videos (Fig.~\ref{fig:PTcompare}).  Analysis throughput is further limited by `tracker fatigue'; in our experience, students/users can rarely process more than 10-20 videos per day without fatigue hindering their decision making.  These challenges have strongly limited particle tracking, preventing it from becoming a widely-used tool in physical and life sciences.  

To tackle these challenges, we developed here a CNN comprised of over 50,000 parameters, and employed machine learning to optimize the network against a diverse array of video conditions.  The end product is a particle tracker (with fully automated identification) that can consistently analyze both 2D and 3D videos with a remarkably low false positive rate, and lower false negative rate, lower localization error and longer average path lengths than a number of the leading particle tracking software.  
The neural network tracker greatly increases the throughput of converting videos into particle position time series, which addresses the biggest bottleneck limiting the applications of particle tracking.

The principal benefit of the statistical nature of the trained CNN is robustness to changing conditions. For example, the net tracker was capable, without any modifications, of tracking salmonella (see Fig.~\ref{fig:frames} far right panel), which are large enough to resolve and appear as rod-shaped in images. Even though the neural net was trained on rotationally symmetric particle shapes, rod-shaped cells were still recognized with strong confidence sufficient for high fidelity tracking. Large polydisperse particles are also readily tracked provided their PSF shape does not deviate too far from the rotationally symmetric training data. Our neural network does not recognize long filaments such as microtubules, which are very far from the targeted particle shapes used in training; such applications will require significant, targeted advances customized to the specific application.  Another example of the robustness of the network is its ability to ignore background objects and effectively suppress false positives. The neural network does not recognize large bright objects that sometimes appear in videos, even though it was trained on images containing slowly varying background intensity. The neural network has also shown remarkable versatility for different applications. While all of the experimental videos used to develop the neural network were of particles suspended in extracellular biological gels, the net tracker has been successfully used to track 30 nm transgenic GFP particles inside living cells and fluorescently-labeled nuclei (data not shown).

The particle localization method utilized the neural network output instead of computing the centroid position from the raw image data (as is typically done), and the resulting localization accuracy was comparable to other methods. However, some applications such as microrheology may require additional accuracy. Several high quality localization algorithms have been developed that potentially might, given a local region of interest (provided by the neural network) in the raw image, estimate the particle center with more accuracy \cite{parthasarathy2012}. One alternative to particle tracking microrheology is differential dynamic microscopy, which uses scattering methods to estimate dynamic parameters from microscopy videos \cite{giavazzi2009scattering}.

Automation opens up new opportunities for 3D particle tracking, harnessing the current wave of advances such as light sheet microscopy.
Visualizing 3D volumetric time series data is a significant challenge.
Most 3D videos contain at least 10-50 times more data than a comparable 2D video. 
Although software-assisted tracking is available for 3D videos, the excessive time needed to verify accurate tracking, coupled with data storage requirements, present significant challanges for broad adoption of 3D tracking. 
By requiring no user-input (for particle identification), we believe the neural network tracker can tackle the longstanding challenge of analyzing 3D videos, and in the process encourage broader adoption of 3D PT. 

Finally, tools based on machine-learning for computer vision are advancing rapidly. 
Applications of neural network-based segmentation to medical imaging are already under development \cite{ronneberger2015u,cciccek20163d,milletari2016v}.
One recent study has used a pixels-to-pixels type CNN to process raw STORM microscopy data into super-resolution images \cite{nehme2018deep}.
The potential for this technology to address outstanding bio-imaging problems is becoming clear, particularly for image segmentation, which is an active research area in machine learning \cite{long2015,zagoruyko2016multipath,pinheiro2016learning,van2016deep,chen2016deeplab,pathak2015constrained,liu2015parsenet}.

\subsection*{Supporting Information (SI)}
\subsubsection*{Simulated videos}

\begin{figure*}[tb]
  \centering
  \includegraphics[width=14cm]{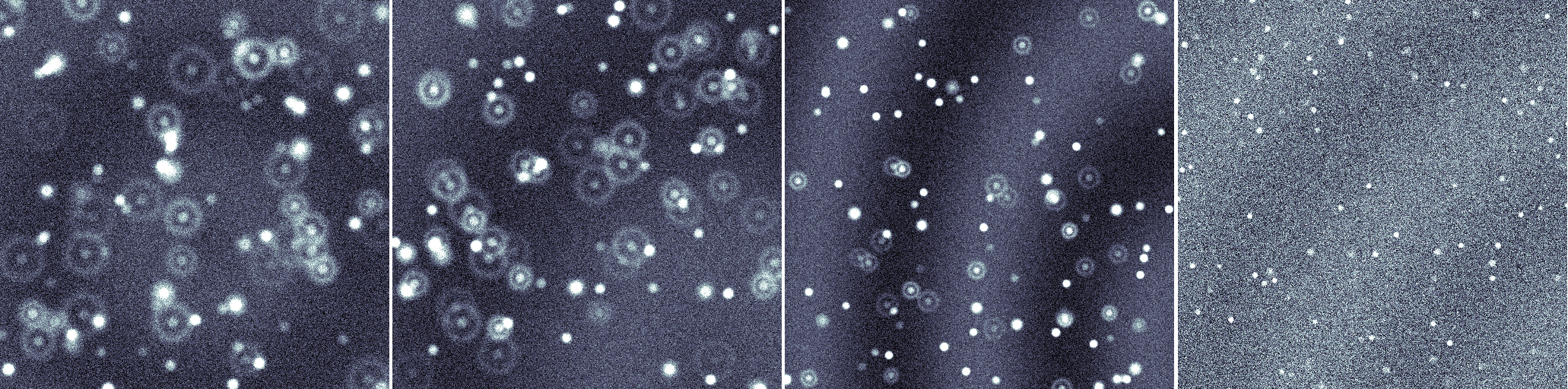}
  \caption{Sample frames from four different synthetic test videos. }
  \label{fig:testFrames}
\end{figure*}

The goal of our simulated videos is to approximate the appearance of real videos for training and testing.
{ \em These videos are not intended to be accurate simulations of particle videos rooted in optical physics.} 
As such, they do not expose physical parameters like wavelength, pixel size, refractive index, numerical apature, etc.
Instead, we postulate a general form that approximates the shape, with a number of parameters that we can use to randomize over a wide range of possible conditions. 
The goal is for the neural network to recognize patterns, independent of the precise details of the optics, particles, and camera used.

Given a particle located at $\xi = (0, 0, 0)$, the observed particle point spread function (PSF) used to generate simulated videos is given by
\begin{multline}
  \psi(x, y, z) = I_1(1 + 0.1(2 h_1 - 1))\left(1 - \gamma \left \vert \tanh(\frac{z}{z_*}) \right \vert \right)
\left \{ 2\exp(-\frac{r^4}{64a^2}) \right . \\
\left.+   (1 - h_2^4)\left[\exp(- \frac{(r - z)^4}{a^4}) + 0.75 h_3 \chi[r<z]\sin ^2 (\left(\frac{\pi r}{z_*}\right)^{3/2})\right] \right\},
\end{multline}
where $r = \sqrt{x^2 + y^2}$.
Here,  $I_1>0$ sets the intensity scale, $z_*$ determines how the PSF fades as the particle moves in $z$, and $a$ determines the PSF radius scale.
The parameters $h_j$, $j=1,2,3$, are values between zero and one, and are intended to randomize the PSF shape and appearance.

A number $N$ of random Brownian particle paths $(x_n(t), y_n(t), z_n(t))$ are generated (using Euler's method) to serve as ground truth.
Then, the image volume at time $t$ is given by
\begin{equation}
  \label{eq:6b}
  I(x, y, z, t) = \sum_{n=1}^N \psi(x - x_{n}(t),y - y_{n}(t), z - z_{n}(t))  + B(x, y, z) + \kappa \Theta(x, y, z, t),
\end{equation}
where $B(x, y, z)$ is a random background intensity, $\kappa>0$ scales the noise,  and $\Theta(x, y, z, t)$ is comprised of i.i.d normal random variables with mean zero and unit variance.
Note that after generating the video using \eqref{eq:6b}, we rounded the output to the nearest integer to more closely represent the integer valued image data most often encountered in experiments.
For randomized background we used
\begin{equation}
  B(x, y) = I_{\rm back} I_1 \sin(\frac{6\pi }{N_x} \sqrt{g_1 (x - g_2 N_x)^2 + g_3(y-g_4N_y)^2}),
\end{equation}
where $I_{\rm back}$ scales the background intensity relative to the PSF and $g_j$ are uniform random variables.

\subsubsection*{Neural network architecture} 
Let the video to be processed by the network be given by $I(x, y, z, t)$, where each dimensional variable is interpreted as indexing discrete pixels (for $x, y$), slices (for $z$), and frames (for $t$).
The video dimensions are $(N_{y}, N_{x}, N_{z}, N_{t})$.

The CNN input is a single image frame from a video:
\begin{equation}
  {\rm Input} =  I(\cdot, \cdot, z, t),
\end{equation}
for fixed $z$-axis slice.
The input is normalized in order to cope a wide range of possible image intensity values.
The image frame $I(\cdot, \cdot, z, t)$ is normalized to have zero mean and unit variance.

\begin{figure*}[htbp]
  \centering
  \includegraphics[width=12cm]{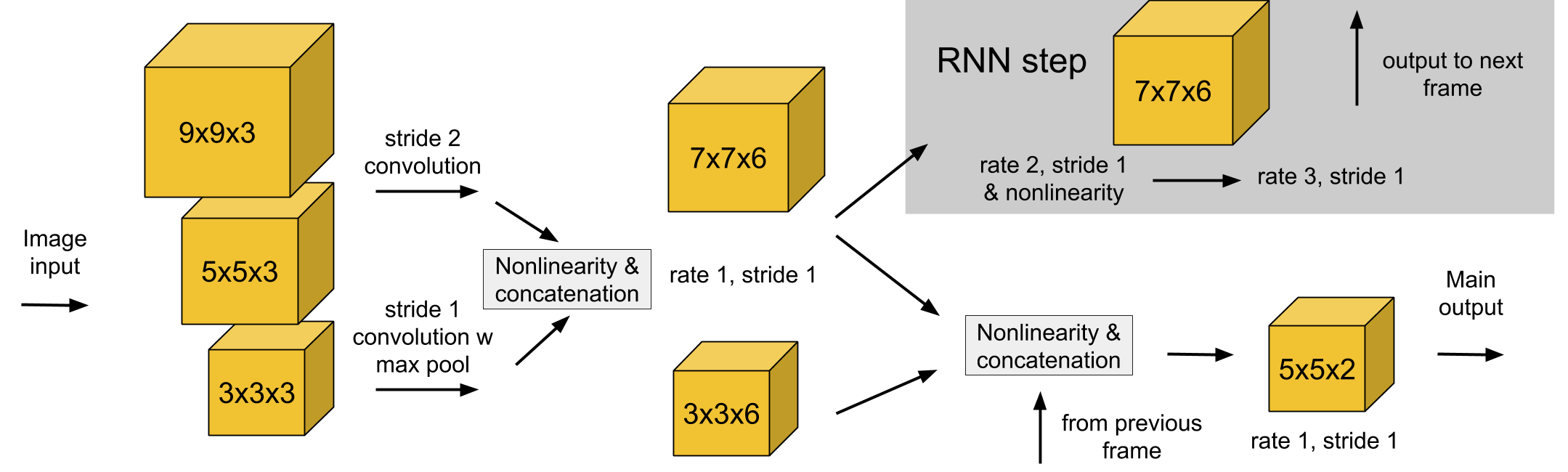}
  \caption{The network architecture of the neural network tracker.}
  \label{fig:archSup}
\end{figure*}

The architecture of the neural network (see Fig.~\ref{fig:archSup}) was designed to manage the number of computationally expensive elements, while maintaining prediction accuracy.
We used the fully-convolutional segmentation network in Ref.~\cite{long2015} as a starting point for our design.
Our first priority was accuracy, followed by evaluation speed.
Evaluation time is largely taken up by convolutions.
Some remaining constraints we considered were training speed, and memory usage.

The CNN is comprised of three convolutional layers and one recurrent layer.
All of the convolution kernels are 4-dimensional arrays whose values are trainable parameters. 
The sizes of the kernels used for each layer are

\begin{itemize}
\item  Layer 1:  $(9, 9, 1, 3)$, $(5, 5, 1, 3)$, and $(3,3,1,3)$ mapping input images to 3 features each (9 total)\\
\item Layer 2:  $(7, 7, 9, 6)$, $(3, 3, 9, 6)$ mapping 9 features to 6 features each (12 total)\\
\item Layer RNN: $(7, 7, 6, 6)$ mapping 6 features to 6 features (this kernel is applied twice successively to the output of kernel 1 of layer 2)\\
\item Layer 3.:  $(5, 5, 18, 2)$ mapping 18 features to the final two output log likelihoods (bilinear interpolation is used to upsample to the original image resolution)
\end{itemize}

The first layer kernels are applied with a stride of two pixels (except for the kernel 3, to which max pooling is applied) so that the layer 1 output has half the $x,y$ resolution as the input (i.e., the layer 2 input tensor has size $(N_{y}/2, N_{x}/2, 9)$). 
In order to maintain a large receptive field with as few trainable parameters as possible, layers 2 and the RNN layer use atrous convolution with a rate of 2 (i.e., they are applied with a stride of 1, but the convolution kernel is applied to a downsampled local patch of the input).
The output of the RNN layer is carried forward to the next frame ($t+1$) and concatenated with the output of layer 2 of frame ($t+1$). The combined 18 features are input into layer 3.
Bilinear interpolation is applied after layer 3 to resample the image to its original resolution.

Nonlinearities are applied after each convolution, using 
\begin{gather}
  {\rm output } = F(\sum_{x',y'}K(x',y'){\rm input}(x+x', y+y') - b), \\
 F(u) = \log(e^{u} + 1).
\end{gather}
Each layer has a separate trainable bias $b$ for each output feature.

Let the output of the interpolation layer be denoted as $L_{n}(x, y)$ for $n=0, 1$.
These outputs are regarded as log likelihoods, at pixel position $x, y$, for background ($n=0$) and the presence of a nearby particle ($n=1$).
The final output of the network is the detection probabilities
\begin{equation}
  p(x, y, z, t) = \frac{e^{L_{1}(x, y)}}{e^{L_{0}(x, y)} + e^{L_{1}(x, y)}},
\end{equation}
Hence, the neural network output, after processing a full video, has the same size and dimension as the video (it may take up more memory since each element is a 32 bit floating point number and videos are typically comprised of 16 bit integers).

\subsubsection*{Neural network training}
Cross entropy is (up to an additive constant that depends on $p$) a measure of how far the approximated distribution $q$ is from the true distribution $p$.
When $q=p$, the cross entropy reduces to the entropy of the true distribution $p$.
Since $p$ never changes for a given training video, our goal is to minimize $H[p, q]$ with respect to $q$ over the entire training set of videos.
At each iteration of the training procedure, a randomly generated training image is processed by the network, the error $H[p, q]$ is computed, and all of the trainable parameters are altered by a small amount (using the gradient decent method explained below) to reduce the observed error.
This training procedure is repeated thousands of times until the error is minimized.

Suppose that all of the trainable parameters are arranged into the vector $\bm{\theta}$. 
The parameters are adjusted at the end of each training iteration $t$ by computing the gradient of $\mathbf{g}_{t} = \nabla_{\theta} H[p_{t}, q_{t}]$.
The gradient vector points in the direction of steepest rate of increase in the error, so the error can be reduced with $\bm{\theta}_{t+1} = \bm{\theta}_{t} - r \mathbf{g}_{t}$, where $r>0$ is a predefined step size.

Generation of training images was performed in Python and training of the neural network was performed using Google's open source software package, Tensorflow \cite{abadi2016tensorflow}.
Training was performed using stochastic gradient descent, with learning rate $0.16$.
The learning rate was decayed exponentially with decay factor $0.95$.
Each iteration of training processed a full $256\times256$ resolution frame from a randomly generated synthetic video, each of which was used for no more than two training iterations.
The training was stopped at 100,000 training iterations.

After training, the neural network is deployed using Tensorflow, which executes the most computationally costly elements of the neural net tracker in highly optimized C++ code. 
Tensorflow can be easily adapted to use multiple cores of a CPU or GPU, depending on available hardware.

\subsubsection*{Parameter values for tracking software used in the synthetic video tests}

Sample frames of the synthetic test videos can be seen in Fig.~\ref{fig:testFrames}.
All of the 2D and 3D videos were tracked using the same set of parameter values.
The neural net tracker uses one parameter in its linking method (for collecting particle localizations into paths). 
The standard deviation for particle displacements was set to $\sigma = 20$.

No method was used with default parameter values. 
Through experimentation, testing 10-15 parameter sets for each method on the full data set, we chose parameter values that showed the best performance overall. 
There was no objectively optimal parameter set since we needed to balance false positives and false negatives.
We chose parameter sets so that the trackers extracted a reasonable fraction of the particle tracks, while maintaining the lowest possible false postive rate.
In practice, paramter values can be tuned to decrease false positives at the expense of fewer extracted tracks.

For Mosaic, we used a custom ImageJ macro to batch process the test videos.
The particle detection parameters were
\begin{equation*}
  {\rm radius} = 8, \quad  {\rm cutoff}  = 0, \quad  {\rm percentile} =0.8
\end{equation*}
For ICY, we used a custom javascript script for batch processing, which only required parameter values for its partical localization method (the particle linking method is fully automated). The particle detection parameters were 
\begin{equation*}
  {\rm scale}_{1} = 0, \quad  {\rm scale}_{2} = 0, \quad  {\rm scale}_{3} = 50 , \quad  {\rm scale}_{4}  = 100
\end{equation*}
For linking, we specified that particles with PSF radius $<2$ (the minimum size in the test videos) be filtered, and that the ICY linker should assume all particles move by standard Brownian motion.

\subsection*{Methods}

HIV, HSV and nanoparticles were prepared as previously described \cite{nunn2015enhanced,wang2014,lai2007rapid}.  Briefly, replication-defective {HIV-1}, internally labeled with an mCherry-Gag construct to avoid alteration of the viral surface, was prepared by transfection of 293T cells with plasmids encoding NL4-3Luc Vpr-Env-, Gag-mCherry, and YU2 Env in a 4:1:1 ratio \cite{nunn2015enhanced}.  Mucoinert nanoparticles were prepared by conjugating 2 kDa of amine-modified polyethylene glycol to carboxyl-modified nanoparticles via a carboxyl-amine reaction; PEG-grafting was verified using the fluorogenic compound 1-pyrenyldiazomethane (PDAM) to quantify residual unmodified carboxyl groups on the nanoparticles \cite{yang2014evading}.  HSV encoding a VP22-GFP tegument protein packaged into HSV-1 at relatively high copy numbers are produced as previously described \cite{wang2014}.  Fluorescent virions or nanoparticles ($\sim$\num{1e8} - \num{1e9} particles per \si{\milli\liter})
 were mixed at 5\% v\v dilution into $\sim$\SI{20}{\micro\liter} of fresh human cervicovaginal mucus collected as previously described \cite{wang2014}, sealed within a custom-made glass chamber.  The translational motions of the particles were recorded using an EMCCD camera (Evolve 512; Photometrics, Tucson, AZ) mounted on an inverted epifluorescence microscope (AxioObserver D1; Zeiss, Thornwood, NY), equipped with an Alpha Plan-Apo 100/1.46 NA objective, environmental (temperature and CO2) control chamber, and an LED light source (Lumencor Light Engine DAPI/GFP/543/623/690). Videos (512x512, 16-bit image depth) were captured with MetaMorph
 
imaging software (Molecular Devices, Sunnyvale, CA) at a temporal resolution of \SI{66.7}{\milli\second} and spatial resolution of \SI{10}{\nano\meter} (nominal pixel resolution \SI{0.156}{\milli\meter} per pixel) for \SI{20}{\second}.  Subpixel tracking resolution was obtained by determining the precise location of the particle centroid by light-intensity-weighted averaging of neighboring pixels. Trajectories were analyzed using ``frame-by-frame'' weighting \cite{wang2015minimizing} in which mean squared displacements (MSD) and effective diffusivities (Deff) are first calculated for individual particle traces. Averages and distributions are then calculated at each frame based on only the particles present in that frame before averaging across all frames in the movie. This approach minimizes bias toward faster-moving particle subpopulations.

\subsection*{Acknowledgements}
Financial support was provided by the National Science Foundation (http://www.nsf.gov) DMS-1715474 (J.M.N), DMS-1412844 (M.G.F.), DMS-1462992 (M.G.F.), and DMR-1151477 (S.K.L.); The David and Lucile Packard Foundation (2013-39274, S.K.L.); and the Eshelman Institute of Innovation (S.K.L).  The funders had no role in study design, data collection and analysis, decision to publish, or preparation of the manuscript. J.M.N would like to thank the Isaac Newton Institute for Mathematical Sciences for support and hospitality during the programme Stochastic Dynamical Systems in Biology when work on this paper was undertaken, including useful discussions with Sam Isaacson, Simon Cotter, David Holcman, and Konstantinos Zygalakis.

\bibliography{refs}
\bibliographystyle{IEEEtran}

\end{document}